\documentclass[a4paper]{article}
\usepackage{epsfig}
\usepackage{amssymb}
\newcommand{\bc}{\begin{center}}
\newcommand{\ec}{\end{center}}
\newcommand{\bd}{\begin{displaymath}}
\newcommand{\ed}{\end{displaymath}}
\newcommand{\be}{\begin{equation}}
\newcommand{\ee}{\end{equation}}
\newcommand{\bea}{\begin{eqnarray}}
\newcommand{\eea}{\end{eqnarray}}
\newcommand{\ba}{\begin{array}}
\newcommand{\ea}{\end{array}}
\newcommand{\bt}{\begin{tabular}}
\newcommand{\et}{\end{tabular}}

\newcommand{\ds}{\displaystyle}

\begin{document}

\hyphenation{MSSM NMSSM Z-boson }

\begin{flushright}
SHEP-05-01\\[8mm]
\end{flushright}

\begin{center}
{\Large \bf  Higgs bosons in the NMSSM with exact and\\[2mm] slightly
broken PQ-symmetry} \\

\vspace{4mm}

D.J.~Miller${}^{1}$, S.~Moretti${}^{2}$, \underline{R.~Nevzorov}${}^{2,3}$\\[2mm] 
\itshape{${}^{1}$ Department of Physics and Astronomy, University of Glasgow, UK}\\
\itshape{${}^{2}$ School of Physics and Astronomy, University of Southampton, UK} \\
\itshape{${}^{3}$ Theory Department, ITEP, Moscow, Russia} \\

\end{center}

\begin{abstract}
\noindent
We explore the Higgs sector of the NMSSM in the limit when the
Peccei--Quinn symmetry is exact or only slightly broken. In this case
the Higgs spectrum has a hierarchical structure which is caused by the
stability of the physical vacuum. We find a strong correlation between
the parameters of the NMSSM if $\kappa=0$ or $\kappa\lesssim
\lambda^2$. It allows one to distinguish the NMSSM with exact or
softly broken PQ-symmetry from the MSSM even when extra scalar and
pseudoscalar Higgs states escape direct detection.
\end{abstract}

\section{Introduction}

Nowadays the simplest supersymmetric (SUSY) extension of the Standard
Model (SM), the so--called Minimal Supersymmetric Standard Model
(MSSM), is possibly the best motivated model beyond the SM. Indeed the
quadratic divergences, that lead to the hierarchy problem in the SM,
are naturally canceled in supersymmetric theories.  By making
supersymmetry local (supergarvity) a partial unification of gauge
interactions with gravity can be achieved. The remarkable coincidence
of gauge coupling constants at the high energy scale $M_X\sim
10^{16}{\rm GeV}$ obtained in the framework of the MSSM allows one to
embed the simplest SUSY extension of the SM into Grand Unified and
superstring theories.

The stabilization of the mass hierarchy in the MSSM does not provide
any explanation for the origin of the electroweak scale, and therefore
a minimal SUSY model should not know about the electroweak scale
before symmetry breaking. However, the MSSM superpotential contains
one bilinear term $\mu (\hat{H}_1 \epsilon \hat{H}_2)$ which is
present before supersymmetry is broken; from dimensional
considerations one would naturally expect the parameter $\mu$ to be
either zero or the Planck scale, but in order to provide the correct
pattern of electroweak symmetry breaking, $\mu$ is required to be of
the order of the electroweak scale. Thus minimal SUSY has its own
``hierarchy'' problem, known as the $\mbox{$\mu$-problem}$.

The most elegant solution of the $\mu$--problem naturally appears in
the framework of superstring--inspired $E_6$ models where the bilinear
terms are forbidden by gauge symmetry. In general these models contain
a few pairs of the Higgs doublets and a few singlet fields
$S_i$. Assuming that only one pair of Higgs doublets and one singlet
survive to low energies the superpotential of the Higgs sector takes
the form $\lambda \hat{S}(\hat{H}_1 \epsilon \hat{H}_2)$. The
considered model includes only one additional singlet field and almost
the same number of parameters as the MSSM. For this reason it can be
regarded as the simplest extension of the MSSM. As a result of
spontaneous symmetry breakdown at the electroweak scale the superfield
$\hat{S}$ gets a non-zero vacuum expectation value ($\langle S \rangle
\equiv s/\sqrt{2}$) and an effective $\mu$-term ($\mu=\lambda
s/\sqrt{2}$) of the required size is automatically generated.

The model discussed above possesses a $SU(2)\times [U(1)]^2$ global
symmetry. Being broken by the vacuum an extended global symmetry leads
to the appearance of a massless CP-odd spinless particle in the Higgs
boson spectrum which is a Peccei-Quinn (PQ) axion. The usual way to
avoid this axion is to introduce a term cubic in the new singlet
superfield $\hat{S}$ in the superpotential that explicitly breaks the
additional $U(1)$ global symmetry. The superpotential of the Higgs
sector of the obtained model, which is the so--called
Next--to--Minimal Supersymmetric Standard Model (NMSSM), is given by
\cite{1}
\be 
W_{H}=\lambda \hat{S}(\hat{H}_1 \epsilon \hat{H}_2)+\frac{1}{3}\kappa\hat{S}^3\,. 
\label{4}
\ee

In this paper we study the Higgs sector of the NMSSM. In Section 2 we
discuss the MSSM limit of the NMSSM.  In Section 3 we investigate the
spectrum and couplings of the Higgs bosons in the exact PQ--symmetry
limit of the NMSSM where $\kappa=0$. The scenario with a slightly
broken PQ--symmetry, where $\kappa$ is small, is considered in Section
4.  The results are summarized in Section 5.

\section{The MSSM limit of the NMSSM}

The Higgs sector of the NMSSM includes two Higgs doublets $H_{1,2}$ and one singlet field $S$. 
The potential energy of the Higgs field interaction can be written as a sum
\be
\ba{rcl}
V&=&V_F+V_D+V_{soft}+\Delta V\,,\\[3mm]
V_F&=&\lambda^2|S|^2(|H_1|^2+|H_2|^2)+\lambda^2|(H_1\epsilon
H_2)|^2+\lambda\kappa\left[S^{*2}(H_1\epsilon
H_2)+h.c.\right]+\kappa^2|S|^4\, ,\\[3mm]
V_D&=&\ds\frac{g^2}{8}\left(H_1^+\sigma_a H_1+H_2^+\sigma_a
H_2\right)^2+\frac{{g'}^2}{8}\left(|H_1|^2-|H_2|^2\right)^2\, ,\\[3mm]
V_{soft}&=&m_1^2|H_1|^2+m_2^2|H_2|^2+m_S^2|S|^2
+\left[\lambda A_{\lambda}S(H_1\epsilon
H_2)+\ds\frac{\kappa}{3}A_{\kappa}S^3+h.c.\right]\, ,\\[2mm]
\ea
\label{5}
\ee
where $H_1^T=(H_1^0,\,H_1^{-})$, $H_2^T=(H_2^{+},\,H_2^{0})$ and
$(H_1\epsilon H_2)=H_2^{+}H_1^{-}-H_2^{0}H_1^{0}$.  At the tree level
the Higgs potential (\ref{5}) is described by the sum of the first
three terms.  $V_F$ and $V_D$ are the $F$ and $D$ terms. Their
structure is fixed by the NMSSM superpotential (\ref{4}) and the
electroweak gauge interactions in the common manner. The last term in
Eq.(\ref{5}), $\Delta V$, corresponds to the contribution of loop
corrections.

The parameters $g$ and $g'$ are $SU(2)$ and $U(1)$ gauge couplings
respectively ($g_1=\sqrt{5/3}g'$), which are known precisely. The
couplings $g,\,g',\,\lambda$ and $\kappa$ do not violate
supersymmetry.  The soft supersymmetry breaking terms are collected in
$V_{soft}$. At the tree level the set of soft SUSY breaking parameters
involves soft masses $m_1^2,\, m_2^2,\, m_S^2$ and trilinear couplings
$A_{\kappa},\, A_{\lambda}$.  The inclusion of loop corrections draw
into the analysis many other soft SUSY breaking terms that define the
masses of different superparticles. All parameters listed above are
here assumed to be real.  In general, complex values of $\lambda$,
$\kappa$ and soft couplings induce CP--violation, and we here restrict
our consideration to the part of the NMSSM parameter space where CP is
conserved.

At the physical minimum of the potential (\ref{5}) the neutral
components of the Higgs doublets $H_1$ and $H_2$ develop vacuum
expectation values $v_1$ and $v_2$ breaking the electroweak symmetry
down to $U(1)$.  Instead of $v_1$ and $v_2$ it is more convenient to
use $\tan\beta=\ds\frac{v_2}{v_1}$ and $v=\sqrt{v_1^2+v_2^2}$, where
$\mbox{$v$=246\,GeV}$ in the physical vacuum.

To start with let us specify the transition from the NMSSM to the
minimal SUSY model which has been studied thoroughly. Because the
strength of the interaction of the extra singlet fields with other
bosons and fermions is determined by the size of the coupling
$\lambda$ in the superpotential (\ref{4}) the MSSM sum rules for the
Higgs masses and couplings are reproduced when $\lambda$ tends to be
zero.  It implies that the vacuum expectation value of the singlet
field should grow with decreasing $\lambda$ as $M_Z/\lambda$ to
ensure the correct breakdown of the electroweak symmetry. The
increasing of $s$ can be achieved either by decreasing $\kappa$ or by
raising $m_S^2$ and $A_{\kappa}$. Since there is no natural reason why
$m_S^2$ and $A_{\kappa}$ should be very large while all other soft
SUSY breaking terms are left in the TeV range, the values of $\lambda$
and $\kappa$ are obliged to go to zero simultaneously so that their
ratio remains unchanged.

Since, in the MSSM limit of the NMSSM, mixing between singlet states
and neutral components of the Higgs doublets vanish, the structures of
the Higgs mass matrices are simplified. This allows one to obtain
simple approximate solutions for the Higgs masses. The NMSSM Higgs
sector involves two charged states with masses
\be
m_{H^{\pm}}^2\approx m_A^2+M_W^2\,,
\label{25}
\ee   
where $M_W=\ds\frac{g}{2}v$ is the mass of the charged W--boson while
\be
m_A^2=\ds\frac{4\mu^2}{\sin^22\beta}\left(x-\frac{\kappa}{2\lambda}\sin2\beta\right)\,,\qquad
x=\frac{1}{2\mu}\left(A_{\lambda}+2\ds\frac{\kappa}{\lambda}\mu\right)\sin 2\beta\,.
\label{251}
\ee
Also there are five neutral fields in the Higgs spectrum. If CP is conserved then
two of them are CP--odd with masses
\be
m_{A_1}^2\approx\ds-3\frac{\kappa}{\lambda}A_{\kappa}\mu\,, \qquad\qquad 
m_{A_2}^2\approx m_A^2\,, 
\label{252}
\ee 
whereas three others are CP--even with masses
\be
\ba{rcl}
m_{h_1}^2 &\approx& \ds 4\frac{\kappa^2}{\lambda^2}\mu^2+\frac{\kappa}{\lambda}A_{\kappa}\mu
+\frac{\lambda^2v^2}{2}x\sin^22\beta-\frac{2\lambda^2v^2\mu^2(1-x)^2}{M_Z^2\cos^22\beta}\, ,\\[3mm]
m_{h_2, h_3}^2 &\approx& \ds\frac{1}{2}\left[m_A^2+M_Z^2\mp \sqrt{(m_A^2+M_Z^2)^2-4m_A^2M_Z^2\cos^2 2\beta}\right]\, ,
\ea
\label{26}
\ee
where $M_Z=\ds\frac{\bar{g}}{2}v$ is a Z--boson mass and
$\bar{g}=\sqrt{g^2+g'^2}$.  In Eqs.(\ref{25})--(\ref{26}) we ignore
the contribution of loop corrections to the Higgs masses.  The terms
of the order of $O(\lambda^2 v^2)$ and $O(\lambda\kappa v^2)$ are also
omitted.  The only exception is for the mass of the CP--even singlet
field $h_1$ where we retain the two last terms proportional to
$\lambda^2 v^2$, since they become significant when $\kappa$ becomes
very small compared to $\lambda$.

For appreciable values of\, $\kappa/\lambda$\, the Higgs spectrum
presented above depends on four variables: $m_A$, $\tan\beta$,
$\ds\frac{\kappa}{\lambda}\,\mu$ and $A_{\kappa}$. The masses of
MSSM--like Higgs bosons ($m_{H^{\pm}}, m_{A_2}, m_{h_2}$ and
$m_{h_3}$) are defined by $m_A$ and $\tan\beta$.  As in the minimal
SUSY model they grow if $m_A$ is increased. At large values of $m_A$
($m_A^2>>M_Z^2$) the masses of the charged, one CP-odd and one CP-even
states are almost degenerate, while the SM-like Higgs boson mass
attains its theoretical upper bound $M_Z |\cos 2\beta|$.  Loop
corrections from the top quark and its superpartners raise this upper
limit up to $130-135\,\mbox{GeV}$. The experimental constraints on the
SUSY parameters obtained in the MSSM remain valid in the the NMSSM
with small $\kappa$ and $\lambda$. For example, non-observation of any
neutral Higgs particle at the LEPII restricts $\tan\beta$ and $m_A$
from below.

The combination of the NMSSM parameters
$\ds\frac{\kappa}{\lambda}\,\mu$ set the mass scale of singlet fields
($m_{h_1}$ and $m_{A_1}$). Decreasing $\kappa$ reduces their masses so
that for $\ds \frac{\kappa}{\lambda} \ll 1$ they can be the lightest
particles in the Higgs boson spectrum.  The parameter $A_{\kappa}$
occurs in the masses of extra scalar and pseudoscalar with opposite
sign, and is therefore responsible for their splitting. Too large a
value of $|A_{\kappa}|$ pulls the mass-squared of either singlet
scalar or singlet pseudoscalar below zero destabilizing the vacuum. An
even stronger lower constraint on $A_{\kappa}$ is found from the
requirement that the vacuum be the global minimum.  Together these
requirements constrain $A_{\kappa}$ and consequently the ratio\,
$(m_{A_1}^2/m_{h_1}^2)$ from below and above
\be
-3\left(\ds\frac{\kappa}{\lambda}\mu\right)^2
\leq \ds A_{\kappa}\cdot\left(\frac{\kappa}{\lambda}\mu\right)\leq 0\, , \qquad\qquad
0\leq \ds \frac{m_{A_1}^2}{m_{h_1}^2}\leq 9\, .
\label{27}
\ee

The main features of the NMSSM Higgs spectrum discussed above are
retained when the couplings $\lambda$ and $\kappa$ increase. In this
case Eqs.(\ref{25})--(\ref{26}) provide some insight into the mass
hierarchy of the NMSSM Higgs sector and qualitatively describe the
dependence of the Higgs masses with respect to the variations of
$m_A,\, \kappa/\lambda,\, A_{\kappa},\, \mu$ and $\tan \beta$.

\section{NMSSM with $\kappa=0$}

The analysis of the MSSM limit of the NMSSM reveals one of the main
impediments in the study of the NMSSM Higgs sector --- the large
number of independent parameters. Indeed even in the limit
$\kappa,\,\lambda\to 0$, when the number of variables parameterizing
the spectrum at the tree level reduces drastically, the masses of the
extra Higgs states take arbitrary values. Therefore it seems very
attractive to take a step back to the simplest extension of the MSSM
when $\kappa=0$.  Since the self interaction of the singlet fields no
longer appears in the superpotential nor in the Higgs effective
potential, there are only 4 parameters defining the masses and
couplings of the Higgs bosons at the tree-level: $\lambda, \mu,
\tan\beta$ and $m_A$ (or $x$). For $\tan\beta\le 2.5$ and small values
of $\lambda$ ($\lesssim 0.1$) the predominant part of the NMSSM
parameter space is excluded by unsuccessful Higgs searches; although
the lightest Higgs boson may partially decouple and not be seen, the
second lightest scalar would be SM-like and visible. Furthermore,
non-observation of charginos at LEPII restricts the effective
$\mu$-term from below: $|\mu|\ge 90-100\,\mbox{GeV}$. Combining these
limits one gets a useful lower bound on $m_A$ at the tree level: 
\be
m_A^2\gtrsim 9M_Z^2 x\,.
\label{28}
\ee

When $\lambda\to 0$ the Higgs boson masses are closely approximated by
Eq.(\ref{25}) and Eqs.(\ref{252})--(\ref{26}) where $\kappa$ and
$A_{\kappa}$ must be taken to be zero. In the considered limit the
mass of the lightest CP--odd Higgs boson, which is predominantly a
singlet pseudoscalar, vanishes. This is a manifestation of the
enlarged $SU(2)\times [U(1)]^2$ global symmetry of the Lagrangian; the
extra $U(1)$ (Peccei--Quinn) symmetry is spontaneously broken giving
rise to a massless Goldstone boson (axion) \cite{10}. The
Peccei--Quinn (PQ) symmetry and its axion allows one to avoid the strong CP
problem, eliminating the $\theta$--term in QCD \cite{11}.

The lightest CP--even Higgs boson is also mainly a singlet field. As
evident from Eq.(\ref{26}), at large values of $\tan\beta$ or $\mu$
the mass--squared of the lightest Higgs scalar becomes negative if the
auxiliary variable $x$ differs too much from unity. Therefore vacuum
stability localizes the auxiliary variable $x$ to a rather narrow range
\be
1-\frac{M_Z|\cos 2\beta|}{m_A^0} < x < 1+\frac{M_Z|\cos 2\beta|}{m_A^0}\, ,
\label{29}
\ee
where $m_A^0=2\mu/\sin 2\beta$. For example, at $\mu=100\,\mbox{GeV}$
and $\tan\beta=3$ the squared mass of the lightest Higgs scalar
remains positive only if $x$ lies between $0.78$ to $1.22$. From the
definition of $m_A$, Eq.(\ref{251}), we see that the tight bounds on
the auxiliary variable $x$ constrain the $m_A$ to the vicinity of
$\mu\, \tan\beta$, which is much larger than the Z-boson mass. As a
result, a mass splitting occurs, where the heaviest CP-odd, CP-even
Higgs and charged Higgs bosons have a mass rather close to $\mu \tan
\beta$, while the SM--like Higgs $h_2$ has a mass of the order of
$M_Z$.

Increasing the value of $\lambda$ increases the lightest Higgs scalar
mass and mixings between the MSSM--like Higgs bosons and singlet
states. As before the masses of the heaviest states are almost
degenerate and close to $m_A\approx\mu\, \tan\beta$.  At the tree level
the masses of the lightest and second lightest Higgs scalars vary
within the limits \cite{4}:
\be
\ba{rcccl}
0 & \le & m_{h_1}^2 & \le & \ds \frac{\lambda^2}{2}v^2x\sin^22\beta\, ,\\[1mm]
\ds M_Z^2\cos^22\beta+\frac{\lambda^2}{2}v^2\sin^22\beta & \le & m_{h_2}^2 & \le &
\ds M_Z^2\cos^22\beta+\frac{\lambda^2}{2}v^2(1+x)\sin^22\beta\, .
\ea
\label{32}
\ee
In Eq.(\ref{32}) the value of $\lambda$ must be smaller than about
$0.7$ to prevent the appearance of Landau pole up below the GUT scale
$M_X$. Although the masses of the lightest Higgs scalars rise with
growing $\lambda$ the mass hierarchy in the Higgs spectrum is
preserved, i.e. $m_{h_1}, m_{h_2}\ll m_{A}$. The couplings of the
lightest CP--even Higgs states to the Z pair ($g_{ZZi}$) and to the
axion and Z--boson ($g_{ZA_1i}$) obey the sum rules \cite{4}
\be
R_{ZZ1}^2+R_{ZZ2}^2\approx 1\, ,
\label{33} 
\ee
\be
R_{ZA_11}^2+R_{ZA_12}^2\approx\ds\frac{1}{4}\left(\frac{\lambda v}{m_A^0}\right)^4\cos^22\beta\,.
\label{330}
\ee
where $R_{ZZi}$ and $R_{ZA_1i}$ are normalized couplings defined by
$g_{ZZi}=\ds\frac{\bar{g}}{2}M_Z\times R_{ZZi}$ and
$g_{ZA_ji}=\ds\frac{\bar{g}}{2}\times R_{ZA_ji}$.

Searches for massless pseudoscalars and light scalar particles at
accelerators as well as the study of their manifestations in
astrophysics and cosmology rule out almost the entire parameter space
of the NMSSM with $\kappa=0$. A particularly stringent constraint
emerges from stellar-evolution \cite{5}--\cite{6}.  Since axions
interact with electrons, nucleons and photons with strength inversely
proportional to the axion decay coupling $f_a\sim s$, they are
produced during the process of star cooling. To evade the modification
of stellar-evolution beyond observational limits one must impose a
lower limit on $f_a$ and $s > 10^9\,\mbox{GeV}$ \cite{7}.

For large values of $\tan\beta$ the restrictions on $s$ are even
stronger. The axion is accompanied by the lightest scalar Higgs boson
(saxion) which has a mass less than $10\,\mbox{KeV}$ for $\tan\beta >
10$ and $\mu< 200\,\mbox{GeV}$, see Eq.(\ref{32}). This light scalar
can be also produced during the cooling of globular--cluster stars
significantly affecting their evolution if the scalar--electron
coupling $g_{Xe}$ is above $1.3\cdot 10^{-14}$ \cite{6}. Since
$g_{h_1e}\sim m_e/s$ this translates into a lower bound on the scale
of PQ--symmetry breaking $f_a\sim s > 10^{11}\,\mbox{GeV}$. Cold dark
matter is composed of both axions and saxions while the supersymmetric
partner to the axion, the axino (the lightest neutralino), is so light
that it does not contribute \cite{L.Hall}.

The constraints on the vacuum expectation value of the singlet field
restrict $\lambda$ to be less then $10^{-6}\,(10^{-9})$ for
$s>10^9\,\mbox{GeV} (10^{11}\,\mbox{GeV})$. The smallness of $\lambda$
could be caused by certain discrete and gauge symmetries which forbid
the operator $\lambda S(H_1H_2)$ at the renormalizable level but which
permit a similar non--renormalizable operator involving additional
singlets, resulting in an effective $\lambda$ proportional to the
ratio of the vacuum expectation values of the new singlet fields and
the Planck scale \cite{King}.  It has also been shown that the
interactions of $S$ with other extra singlet fields result in the
stabilization of the Higgs scalar potential which otherwise has a
direction unbounded from below when $\kappa=0$ and
$m_S^2<0$. Moreover, this axion could play the role of an inflaton
field \cite{King1}.

For tiny values of $\lambda$, the decoupling of new singlet states
prevents their observation at future colliders.  Thus if the NMSSM
with unbroken global PQ--symmetry is realized in nature, only
MSSM--like Higgs bosons will be discovered in the near
future. However, the strong correlation between the masses of the
heaviest Higgs bosons and $\mu \tan \beta$ revealed in the tree level
analysis (see Eq.(\ref{29})) does not take place in the MSSM but must
be observed here(see also \cite{131}).  The inclusion of loop
corrections does not change the qualitative pattern of the Higgs
spectrum and does not enlarge the allowed range of $x$ (or
$m_A$). Loop corrections only slightly shift the admissible range of
the variable $x$ which shrinks with increasing $\mu$ and $\tan\,
\beta$ in compliance with Eq.(\ref{29}).  In order to demonstrate the
correlation between $m_A$ and $\mu \tan\beta$ we examined $10^6$
different scenarios, with $m_A$ and $\tan \beta$ chosen randomly
between $0$ to $6$~TeV and $3$ to $30$ respectively. We calculated the
one-loop mass spectrum and, for every scenario with a stable vacuum,
plotted a single point on the $m_A$--$\tan\beta$ plane of
Fig.1. We discarded scenarios with unstable
vacua. It is immediately evident that the physically acceptable
scenarios all lie within a small area around $m_A \approx \mu \tan
\beta$ \cite{7}. Since the positivity or negativity of $m_{h_1}^2$ is
independent of $s$ (or $\lambda$), the NMSSM with $\kappa=0$ is ruled
out for all large values of the singlet expectation value if after
measuring $\mu$ and $\tan\beta$ at future accelerators, the heavy
pseudoscalar mass is not found to lie close to $\mu \tan \beta$.
Alternatively, if the mass prediction were found to hold, it would
provide an indirect evidence for the PQ--symmetric NMSSM as a solution
to the strong CP problem and for the axion and saxion as a source of
dark matter.
\begin{figure}[h]
\label{fig:matbscan}
\begin{center}
{\hspace*{-0mm}\includegraphics[scale=0.6]{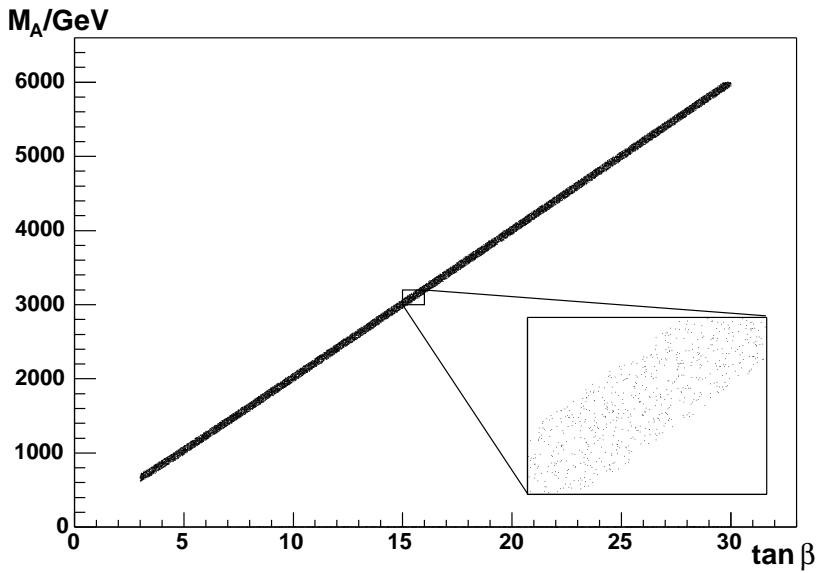}}\\
\caption{\it The distribution of scenarios with physically acceptable
vacua, with $M_A$ chosen randomly between $0$ and $6$~TeV, $\tan\beta$
chosen randomly between $3$ and $30$ and $\mu=200\,\mbox{GeV}$.  The
blow-up allows individual scenario points to be seen.}
\end{center}
\end{figure}

\section{Slight breaking of the PQ--symmetry}

If one wants to avoid the introduction of a new intermediate scale
that arises in the NMSSM with $\kappa=0$ when the astrophysical limits
on the couplings of the lightest scalar and pseudoscalar particles are
applied, one must break the Peccei--Quinn symmetry. For a discussion
of the possible origins of this symmetry breaking in the
NMSSM see Refs.\cite{131}-\cite{13}. Here we assume that the
violation of the Peccei--Quinn symmetry is caused by a non--zero value
of $\kappa$.  Moreover we restrict our consideration to small values
of $\kappa$ when the PQ--symmetry is only slightly broken. To be
precise we consider values of $\kappa$ that do not greatly change the
vacuum energy density:
\be \kappa \lesssim \lambda^2\,.
\label{34}
\ee
If $\kappa \gg \lambda^2$ then the terms $\kappa^2|S|^4$ and
$\ds\frac{\kappa}{3}A_{\kappa}S^3$ in the Higgs effective potential
(\ref{5}) become much larger than $|\mu|^4\sim M_Z^4$ increasing the
absolute value of the vacuum energy density significantly.

For small values of $\lambda$ the approximate formulae
Eqs.(\ref{252})--(\ref{26}) obtained in section~2 remain valid.
However, breaking the PQ--symmetry gives the lightest CP--odd Higgs an
appreciable mass that allows it to escape the strong astrophysical
constraints previously outlined. One must ensure that the value of
$\kappa$ is large enough for the lightest scalar and pseudoscalar to
escape the exclusion limits of LEP, but it is still physically
reasonable to only slightly break the PQ--symmetry, as defined by
Eq.(\ref{34}).

Indeed, for the appreciable values of $\kappa$ and $\lambda$ this slight
breaking of the Peccei--Quinn symmetry may arise naturally from their
renormalization group (RG) flow from $M_X$ to $M_Z$ \cite{14}.  While
the values of the Yukawa couplings at the Grand Unification scale
grow, the region where the solutions of the RG equations are
concentrated at the electroweak scale shrinks and they are focused
near the quasi fixed point \cite{15}:
\be
h_t(M_t)\approx1.103,\qquad \lambda(M_t)\approx 0.514,\qquad \kappa(M_t)\approx 0.359\, .
\label{35}
\ee
This point appears as a result of intersection of the Hill-type
effective surface with the invariant line that connects the stable
fixed point in the strong Yukawa coupling limit with the infrared
fixed point of the NMSSM renormalization group equations. The
requirement of perturbativity up to the Grand Unification scale
provides stringent restrictions on the values of $\lambda(M_t)$ and
$\kappa(M_t)$
\be
\lambda^2(M_t)+\kappa^2(M_t)< 0.5\, .
\label{34}
\ee

In order to obtain a realistic spectrum, one must of course include
the leading one--loop corrections from the top and stop loops. We
performed this exercise numerically \cite{14} and present in Fig.2
the mass spectrum as a function of $m_A$ for the parameters
$\lambda=0.3$, $\kappa=0.1$, $\mu=157$\,GeV, $\tan \beta=3$ and
$A_{\kappa}=-60\,\mbox{GeV}$.
\begin{figure}[h]
\label{fig:masses}
\begin{center}
{\hspace*{-0mm}\includegraphics[totalheight=70mm,keepaspectratio=true]{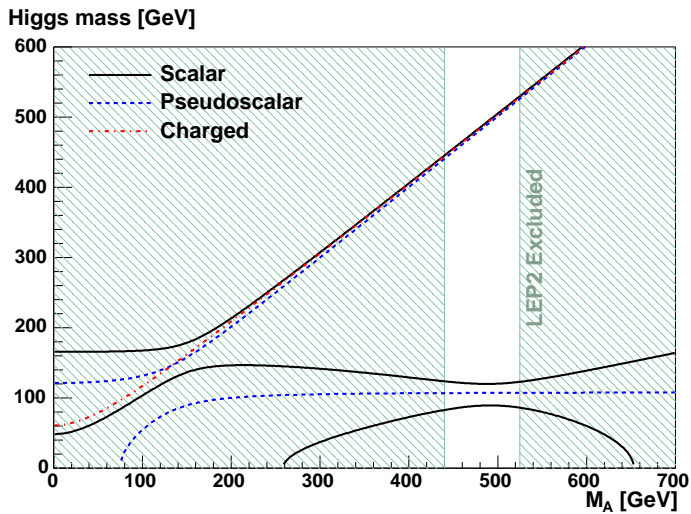}}
\end{center}
\caption{\it The dependence of the Higgs boson masses on $m_A$ for
$\lambda=0.3$, $\kappa=0.1$, $\mu=157$\,GeV, $\tan \beta=3$ and
$A_{\kappa}=-60\,\mbox{GeV}$. Solid, dashed and dashed--dotted curves
correspond to the one--loop masses of the CP-even, CP-odd and charged
Higgs bosons respectively. All masses are given in GeV.}
\end{figure}
One sees that most of the structure outlined above is retained. The
heaviest scalar and pseudoscalar are approximately degenerate with the
charged Higgs boson and track $m_A$. The second lightest scalar is of
the order of the $Z$-boson mass plus radiative corrections, mimicking
the lightest scalar of the MSSM. The breaking of the PQ--symmetry has
raised the masses of the lightest scalar and pseudoscalar to values
which agree very well with the approximate expression of
Eqs.(\ref{252})-(\ref{26})\footnote{The agreement with tree-level
expressions is good because the singlet nature of the new fields
suppresses loops corrections.}. Also notice that the vacuum stability
prevents having very high or very low values of $m_A$ (or $x$) but now
the allowed range has increased significantly, permitting $m_A$ to
substantially deviate from $\mu \tan \beta$.

One might expect that such a light Higgs scalar should already be
ruled out by LEP, but this is not the case \cite{16}. The reduced
coupling to the $Z$-boson allows for a singlet like scalar
substantially below the current SM LEP bounds. Indeed, LEP limits have
been included in Fig.2 as a shaded area: for this parameter choice,
values of $m_A$ in the shaded region either provide a scalar Higgs
boson which would have been seen at LEP or have an unstable
vacuum. There is a substantial range in $m_A$, once more around the
value $m_A \approx \mu \tan \beta$, where the Higgs scalar remains
undetected. In this way, the mass hierarchy between the lighter Higgs
bosons, around the electroweak scale, and the heavier Higgs bosons, at
around $\mu \tan \beta$, is maintained. Since the coupling of the
lightest scalar to the $Z$-boson must necessarily be suppressed in
this region to avoid detection at LEP, the sum rule of Eq.(\ref{33})
tells us that the couplings of the second lightest scalar will be
similar to those of the lightest scalar in the MSSM. It is interesting
to note that this light scalar would be very difficult to see at the
LHC since it will principally decay hadronically, presenting a signal
which is swamped by huge QCD backgrounds. According to Eq.(\ref{330}),
the coupling of the pseudoscalar to the $Z$-boson is always rather
small. Nevertheless, if these light Higgs bosons could be seen, one
would have a definitive signature of the NMSSM, even without observing
the heavier states.

\section{Conclusions}

We have studied the Higgs sector of the NMSSM with exact and slightly
broken Peccei--Quinn symmetry.  In the PQ--symmetric NMSSM
astrophysical observations exclude any choice of the parameters unless
one allows $s$ to be enormously large
$(>10^9-10^{11}\,\mbox{GeV})$. These huge vacuum expectation values of
the singlet field can be consistent with the electroweak symmetry
breaking only if the coupling $\lambda$ is extremely small
$10^{-6}-10^{-9}$. Such tiny values of $\lambda$ may arise from
non--renormalizable operators. In this limit the main contribution to
the cold dark matter density comes from axion and saxion contributions
while that of the lightest supersymmetric particle, the axino, is
negligible.

If the PQ--symmetry is exact or only slightly broken, vacuum stability
and LEP exclusion require parameters which cause a splitting in the
NMSSM Higgs spectrum. The heaviest scalar, heaviest pseudoscalar and
charged Higgs bosons are approximately degenerate with masses around
$m_A\approx\mu\tan\beta$. The other three neutral states are
considerably lighter. The masses of the lightest scalar and
pseudoscalar, which are predominantly singlet fields, are governed by
the combination of parameters $\ds\frac{\kappa}{\lambda}\,\mu$.  The
SM--like Higgs boson mass remains at the electroweak scale.

In the limit of vanishing $\lambda$ and $\kappa$ the extra CP--even
and CP--odd singlet states decouple from the rest of the spectrum and
become invisible. However in the case of exact PQ--symmetry with $\kappa=0$ (or
very slightly broken PQ--symmetry with $\kappa\ll \lambda^2$) the stability of
the physical vacuum constrains the masses of the heavy Higgs bosons in
the vicinity of $m_A\approx \mu\, \tan\beta$. The strong correlation
between $m_A$ and $\mu \tan\beta$ coming from the dark sector of the
NMSSM gives a unique ``smoking gun'' for distinguishing this model
from the MSSM even if no extra Higgs states are discovered.

For appreciable values of $\lambda$ and $\kappa$ the slight breaking
of the PQ--symmetry can be caused by the NMSSM renormalization group
flow. Increasing $\lambda$ increases the mixing between the light
CP--even Higgs bosons, while increasing $\kappa$ increases the masses
of the predominantly singlet states. For small values of $\kappa$, one
can have a light scalar Higgs boson which would not have been seen at
LEP. Although the range of $m_A$ allowed by vacuum stability increases
significantly, one is still required to have $m_A \approx \mu \tan
\beta$ in order to avoid the LEP constraints, leading to a mass
splitting between the light and heavy Higgs bosons.  Observing two
light scalars and one pseudoscalar Higgs but no charged Higgs boson at
future colliders would yield another opportunity to differentiate the
NMSSM with a slightly broken PQ--symmetry from the MSSM even if the
heavy Higgs states are inaccessible.

\section*{Acknowledgements}
\noindent
RN is grateful to E.Boos, I.Ginzburg, S.King, M.Krawczyk and
C.Panagiotakopoulos for fruitful discussions and helpful remarks.  The
work of RN was partly supported by the Russian Foundation for Basic
Research (projects 00-15-96562 and 02-02-17379) and by a Grant of
President of Russia for young scientists (MK--3702.2004.2).



\begin{thebibliography}{99}

\bibitem{1}
P.Fayet, Nucl.Phys. B 90 (1975) 104;
H.P.Nilles, M.Srednicki, D.Wyler, Phys.Lett.B 120 (1983) 346;
J.M.Frere, D.R.T.Jones, S.Raby, Nucl.Phys.B 222 (1983) 11;
J.P.Derendinger, C.A.Savoy, Nucl.Phys.B 237 (1984) 307.
M.I.Vysotsky, K.A.Ter-Martirosian, Sov.Phys.JETP 63 (1986) 489;
J.Ellis, J.F.Gunion, H.Haber, L.Roszkowski, F.Zwirner, Phys.Rev.D 39
(1989) 844.
L.Durand, J.L.Lopez, Phys.Lett.B 217 (1989) 463;
L.Drees, Int.J.Mod.Phys.A 4 (1989) 3635.
\bibitem{10}
S.Weinberg, Phys.Rev.Lett. 40 (1978) 223; F.Wilczek, Phys.Rev.Lett. 40 (1978) 279.
\bibitem{11}
R.D.Peccei, H.R.Quinn, Phys.Rev.Lett. 38 (1977) 1440; Phys.Rev.D 16 (1977) 1791.
\bibitem{4} R.Nevzorov, D.J. Miller, {\it Proceedings to the 7th
Workshop "What comes beyond the Standard Model"}, ed. by
N.S.Mankoc-Borstnik, H.B.Nielsen, C.D.Froggatt, D.Lukman,
DMFA--Zaloznistvo, Ljubljana, 2004, p.107; hep-ph/0411275.
\bibitem{5}
D.A.Dicus, E.W.Kolb, V.L.Teplitz and R.V.Wagoner, Phys.Rev.D 18 (1978) 1829;
G.Raffelt, A.Weiss, Phys.Rev.D 51 (1995) 1495.
\bibitem{6}
J.A.Grifols and E.Masso, Phys.Lett.B 173 (1986) 237;
J.A.Grifols, E.Masso, S.Peris, Mod.Phys.Lett.A 4 (1989) 311;
\bibitem{7}
D.J.Miller, R.Nevzorov, hep-ph/0309143.
\bibitem{L.Hall} 
B.Feldstein, L.J.Hall, T.Watari, hep-ph/0411013.
\bibitem{King}
M.Bastero-Gil, S.F.King, Phys.Lett.B 423 (1998) 27.
\bibitem{King1}
O.J.Eyton-Williams,S.F.King, hep-ph/0411170.
\bibitem{131}
C.Panagiotakopoulos, A.Pilaftsis, Phys.Rev.D 63 (2001) 055003.
\bibitem{13}
C.Panagiotakopoulos, K.Tamvakis, Phys.Lett.B 469 (1999) 145;
R.B.Nevzorov, M.A.Trusov,J.Exp.Theor.Phys.91 (2000) 1079;
A.Dedes, C.Hugonie, S.Moretti, K.Tamvakis, Phys.Rev.D 63 (2001) 055009;
R.B.Nevzorov, K.A.Ter-Martirosyan, M.A.Trusov, Phys.Atom.Nucl.65 (2002) 285.
\bibitem{15}
R.B.Nevzorov, M.A.Trusov, Phys.Atom.Nucl.64 (2001) 1299.
\bibitem{14}
D.J.Miller, R.Nevzorov, P.M.Zerwas, Nucl.Phys.B 681 (2004) 3.
\bibitem{16} D.J.Miller, S.Moretti, {\it Physics interplay of the LHC
and the ILC}, ed by G.Weiglein, T.Barklow, E.Boos et al., The LHC/LC
Study Group, 2004, p.124; hep-ph/0403137.
                                                                                                                                                  
\end{thebibliography}
\end{document}